\documentstyle[aps,epsf]{revtex}
\def\COBE{{\rm COBE}}
\def\df{{\rm d}}
\def\e{{\rm e}}
\def\GeV{{\rm GeV}}
\def\H{{\rm H}}
\def\inf{{\rm inf}}
\def\SUSY{{\rm SUSY}}
\def\therm{{\rm therm}}
\begin{document}
\preprint{\small OUTP-99-39P}
\title{Implementing quadratic supergravity inflation}
\author{\large Gabriel Germ\'an,\footnote{On leave from Centro de Ciencias
                F\'{\i}sicas, Universidad Nacional Aut\'onoma de M\'exico,
                62250 Cuernavaca, Morelos, M\'exico}
               Graham Ross and
               Subir Sarkar \medskip}
\address{Theoretical Physics, University of Oxford,
         1 Keble Road, Oxford OX1 3NP, UK}
\date{\today}
\maketitle
\begin{abstract}
We study inflation driven by a slow-rolling inflaton field,
characterised by a quadratic potential, and incorporating radiative
corrections within the context of supergravity. In this model the
energy scale of inflation is not overly constrained by the requirement
of generating the observed level of density fluctuations and can have
a physically interesting value, e.g. the supersymmetry breaking scale
of $10^{10}$ GeV or the electroweak scale of $10^3$ GeV. In this mass
range the inflaton is light enough to be confined at the origin by
thermal effects, naturally generating the initial conditions for a
(last) stage of inflation of the new inflationary type.
\end{abstract}

\section{Introduction}\label{intro}

The modern paradigm for cosmological inflation is generically termed
`chaotic', referring primarily to its choice of initial conditions
\cite {book}. In its simplest form, a single scalar field --- the
inflaton $\phi$ --- evolves slowly along its potential $V(\phi)$
towards its minimum at the origin. The essential challenge is to embed
consistently this potential in an underlying particle physics model,
within a natural range of variation of the parameters involved and
respecting the `slow-roll conditions' \cite{review}. These are upper
limits on the normalized slope and curvature of the potential:
\begin{equation}
 \epsilon \equiv \frac{M^2}{2}\left(\frac{V'}{V}\right)^2 \ll 1 ,\qquad
 |\eta| \equiv M^2\left|\frac{V''}{V}\right| \ll 1 .
\label{slowroll}
\end{equation}
Here $M\,(\equiv\,M_{\rm P}/\sqrt{8\pi}\simeq2.44\times10^{18}\GeV)$
is the normalised Planck mass and the potential determines the Hubble
parameter during inflation as,
$H_\inf\equiv\dot{a}/a\simeq\sqrt{V/3M^2}$.  Inflation ends
(i.e. $\ddot{a}$, the acceleration of the cosmological scale factor,
changes sign from negative to positive) when $\epsilon$ and/or
$|\eta|$ become of ${\cal O}(1)$.

It is easy to see that this challenge is {\em not} met by simple
models of the potential. Consider the quadratic form \cite{chaotic}
\begin{equation}
 V (\phi) = \frac{1}{2} m_\phi^2 \phi^2 .
\label{mono}
\end{equation}
The adiabatic scalar density perturbation generated through quantum
fluctuations of the inflaton is \cite{review}
\begin{equation}
 \delta^2_\H (k) = \frac{1}{150\pi^2} \frac{V_\H}{M^4}
                   \frac{1}{\epsilon_\H}\ ,
\label{deltah}
\end{equation}
where the subscript $\H$ denotes the epoch at which a fluctuation of
wavenumber $k$ crosses the Hubble radius $H^{-1}$ during inflation,
i.e. when $aH=k$. (We normalise $a=1$ at the present epoch, when the
Hubble expansion rate is $H_0\equiv100h$~km\,s$^{-1}$Mpc$^{-1}$, with
$h\sim0.5-0.8$.) The COBE observations \cite{cobe} of anisotropy in
the cosmic microwave background on large angular-scales require
\cite{review}
\begin{equation}
 \delta_\COBE \simeq 1.9 \times 10^{-5} ,
\label{cobe}
\end{equation}
on the scale of the observable universe
($k_\COBE^{-1}\sim\,H_0^{-1}\sim3000h^{-1}$~Mpc). In addition, the
COBE data fix the spectral index,
$n_\H(k)\equiv1+\df\delta_\H^2(k)/\df\ln\,k=1-6\epsilon_\H+2\eta_\H$,
on this scale:
\begin{equation}
 n_\COBE = 1.2 \pm 0.3 .  \label{ncobe}
\end{equation}
The slow-roll condition (\ref{slowroll}) together with the COBE
normalisation (\ref{cobe}) then implies
\begin{equation}
 \langle\phi\rangle \gg M, \quad m_\phi \ll H_\inf \ll 10^{15} \GeV ,
\label{vev}
\end{equation}
at the epoch when the fluctuations observed by COBE were generated.
Similarly if we consider the quartic potential $V=\case{1}{4}\lambda\phi^4$
\cite{chaotic}, eqs.(\ref{slowroll}) and (\ref{cobe}) imply
\begin{equation}
 \langle\phi\rangle \gg M, \qquad \lambda \ll 10^{-11} .
\label{lambda}
\end{equation}

Thus we see that the slow-roll condition (especially on $\eta$)
generically requires $\langle\phi\rangle\gg\,M$ which is difficult to
motivate in a particle physics model. In models incorporating gravity,
non-renormalisable corrections of the form $(\langle\phi\rangle/M)^n$
are unsuppressed at large $\langle\phi\rangle$ for any $n$ and do not
have the monomial form usually assumed in chaotic inflationary
models. Furthermore the COBE measurement of the scalar perturbation
amplitude implies extreme fine tuning because the only natural scale
for $m_\phi$ is the Planck scale $M$, in conflict with
eq.(\ref{vev}).\footnote{This is just the generic mass `hierarchy
problem' associated with fundamental scalar fields.} Alternatively an
extremely small value is required for the self-coupling as in
eq.(\ref{lambda}), which also represents unnatural fine tuning.

Given the difficulty of constructing realistic particle physics models
which can drive inflation for large $\langle\phi\rangle$, it is
natural to ask whether there are viable models which work for small
$\langle\phi\rangle$, i.e. `new inflation' \cite{new} in which the
potential has a maximum at the origin and the inflaton evolves away
from it. In this case low powers of $\langle\phi\rangle/M$ will
dominate the potential during the era of observable
inflation. Starting from $\langle\phi\rangle\sim0$ and assuming that
the symmetry properties of the model forbid a linear term, the
quadratic term will dominate, giving a potential of the form
\begin{equation}
 V (\phi) \sim \Delta^4 - \frac{1}{2} m_\phi^2 \phi^2 + \dots ,
\label{newinflation}
\end{equation}
where the constant vacuum energy $\Delta^4$ is now the leading term in the
potential. From eq.(\ref{slowroll}) we have now the constraint
\begin{equation}
 m_\phi \ll H_\inf \sim \Delta^{2}/M ,
\label{massbound}
\end{equation}
and this is again much smaller than the natural value for the mass. To
solve this problem we are driven to consider theories in which a
symmetry prevents $m_\phi$ from being large. The only known symmetry
capable of achieving this is supersymmetry which can guarantee that
$m_\phi$ vanishes in the limit that supersymmetry is unbroken. However
the non-vanishing potential of eq.(\ref{newinflation}) driving
inflation {\em breaks} (global) supersymmetry and so, even in
supersymmetric models, $m_\phi$ will be non-zero in general. In the
extreme case that the inflaton has vanishing non-gravitational
couplings, gravitational effects will typically induce a mass of order
$H_\inf\sim\Delta^2/M$ for any scalar field \cite{susybreak}, in
particular the inflaton \cite{copeland}. Nevertheless this is a big
improvement over the non-supersymmetric case, for the fine tuning
problem now simply becomes one of requiring
$m_\phi=\beta\Delta^2/M$ with $\beta\lesssim0.1$ to obtain
successful inflation.

In this Letter we summarise the results of a general analysis of
supergravity inflationary models in which a quadratic term dominates
the evolution of the potential for small values of the inflaton. As
discussed in Section~\ref{eta}, one can find models for which $\beta$
is naturally small, consistent with the slow-roll
requirements. However, even in these models it is difficult to make
the mass negligibly small so that the quadratic term is likely to be
dominant at small inflaton field values. For this reason we consider
models of inflation driven by a quadratic term to be the most {\em
natural} supersymmetric inflationary models. However such models still
leave unanswered several important questions concerning the initial
conditions.

The most important question is why the universe should initially be
homogeneous enough for slow-roll inflation to begin \cite{book}. In
chaotic models \cite{chaotic}, inflation begins when the scale of the
potential energy is of order the Planck scale and the horizon (the
scale over which the universe must be homogeneous for inflation to
start) is also of order the Planck scale. By contrast in the models
discussed here, inflation starts much later when the horizon contains
many such Planck scale horizons and, in this case, it is difficult to
understand how the necessary level of homogeneity can be
realised. However this argument is not really a criticism of the
possibility that there be a late stage of inflation but rather a
statement that this cannot be all there is. Thus we are implicitly
assuming that there was some other process which ensured the necessary
homogeneity at the begining of quadratic inflation, as in the original
chaotic inflation scenario \cite{chaotic} or in the context of quantum
cosmology \cite{hawking}. Such models lead to a situation in which a
homogeneous universe emerges at the Planck era and potential energy is
released reheating the universe and possibly setting the conditions
for further periods of inflation to occur.

This leads to the second initial condition question, namely why should
such a universe subsequently undergo inflation at a lower energy
scale? In the case of interest here this amounts to asking why the
inflaton field has initially such a small value that slow-roll
inflation from the neighbourhood of the origin is possible?
Remarkably, within the context of supersymmetric inflationary models
there is a straightforward answer because quite generally
supersymmetry is broken by finite temperature effects i.e.  the
inflaton potential at high temperatures is not likely to be the same
as that at zero temperature
\begin{equation}
 V (\phi ,T) \left|_{\min} \neq V (\phi, 0) \right|_{\min}
\end{equation}
As we discuss later, thermal effects may readily drive the inflaton to
the origin where the symmetry is enhanced, thus offering an elegant
solution to this initial condition problem. Of course this requires
that the system be initially in thermal equilibrium and this is not
normally the case in slow-roll inflationary models, particularly since
the inflaton should be very weakly coupled in order not to spoil the
required flatness of its potential. However quadratic inflation is
special in as much as the value of the potential during inflation is
not strongly constrained by the need to generate the correct magnitude
of density fluctuations. For the case that the potential energy
driving inflation is low, we will show that the processes leading to
thermal equilibrium have time to work before the inflationary era
starts.

A further possible advantage of lowering the inflationary scale is
that it may be then readily identified with a scale already present in
particle physics. For example a low $\Delta$ may be identified with
the SUSY breaking scale in the hidden sector,
$\Delta_\SUSY\sim10^{10}\GeV$. As is well known \cite{bailin} this
yields a gravitino mass,
\begin{equation}
 m_{3/2} \sim \frac{\Delta_\SUSY^2}{M}\sim 10^2 \GeV,
\end{equation}
of order the electroweak scale, as is needed to avoid the hierarchy
problem of Grand Unified Theories. Recently an even more radical
solution to the hierarchy problem has been suggested, namely that the
only fundamental scale is the electroweak breaking scale itself and
that the (four dimensional) Planck scale is a derived quantity
\cite{largedim}. As we shall see, slow-roll quadratic inflation is
possible even at the electroweak scale. In the next Section we discuss
the expectation for the inflaton potential following from
supergravity. Readers primarily interested in the cosmological
implications may wish to skip this and go straight to Section~\ref{para}.

\section{The Supergravity potential} \label{eta}

In $N=1$ supersymmetric theories with a single SUSY generator,
complex scalar fields are the lowest components, $\phi^a$, of chiral
superfields, $\Phi^a$, which contain chiral fermions, $\psi^a$, as
their other components. In what follows we will take $\Phi^a$ to be
left-handed chiral superfields so that $\psi^a$ are left-handed massless
fermions. Masses for fields will be generated by spontaneous symmetry
breakdown so that the only fundamental mass scale is the normalised Planck
scale. This is aesthetically attractive and is also what follows if the
underlying theory generating the effective low-energy supergravity theory
emerges from the superstring. The $N=1$ supergravity theory describing the
interaction of the chiral superfields is specified by the K\"{a}hler
potential \cite{bailin},
\begin{equation}
 G (\Phi, \Phi^\dagger) = d (\Phi, \Phi^\dagger) + \ln |f(\Phi)|^2 .
\label{g}
\end{equation}
Here $d$ and $f$ (the superpotential) are two functions which need to be
specified; they must be chosen to be invariant under the symmetries of the
theory. The dimension of $d$ is 2 and that of $f$ is 3, so terms bilinear
(trilinear) in the superfields appear without any mass factors in $d$ ($f$).
The scalar potential following from eq.(\ref{g}) is given by \cite{bailin}
\begin{equation}
 V = \exp\left(\frac{d}{M^2}\right)
     \left[F^{A\dagger}(d_A^B)^{-1}F_{B} -
     3\frac{|f|^2}{M^2}\right] + {\rm D-terms} ,
\label{V}
\end{equation}
where
\begin{equation}
 F_{A} \equiv \frac{\partial f}{\partial \Phi^A} +
       \left(\frac{\partial d}{\partial\Phi^A}\right) \frac{f}{M^2} ,\qquad
 \left(d_A^B\right)^{-1} \equiv
 \left(\frac{\partial^2
d}{\partial\Phi^A\partial\Phi_B^\dagger}\right)^{-1}.
\end{equation}
At any point in the space of scalar fields $\Phi$ we can make a combination
of a K\"{a}hler transformation and a holomorphic field redefinition such
that $\phi^a=0$ at that point and the K\"{a}hler potential takes the form
$d=\sum_{a}|\Phi_a|^2+\ldots$. In this form, the scalar kinetic terms
are canonical at $\phi^a=0$ and from eq.(\ref{V}), neglecting D-terms and
simplifying to the case of a single scalar field, the scalar potential
writes
\begin{equation}
 V = \left(\e^{|\phi|^2/M^2 + \dots}\right)
  \left[
    |(f_\phi + f\phi^\ast + \dots)(1 + \dots)|^2 
    - 3 \frac{|f|^2}{M^2} \right] 
   = V_{0} + |\phi|^2 \frac{V_0}{M^2} + \dots
\label{sugra}
\end{equation}
where $V_{0}\equiv\,V|_{\phi=0}$. Since inflation is driven by the
non-zero value of $V$, we see that the resultant breaking of
supersymmetry gives all scalar fields a contribution to their
mass-squared of $V_0/M^2$, in conflict with the slow-roll condition
(\ref{slowroll}) on $\eta$. This is the essential problem one must
solve if one is to implement inflation in a supergravity
theory. However, as stressed earlier, the problem is relatively mild
when compared to the non-supersymmetric case because the suppression
for $\eta$ need only be by a factor of 10 or so.

\subsection{Solving the $\eta$ problem}

There have been several proposals for dealing with this problem. One
widely explored possibility is D-term inflation \cite{dterm}. In
particular one may consider an anomalous D-term in eq.(\ref{V}) of the
form
\begin{equation}
 \frac{g^2}{2}(\xi -\sum_i q_i |\widetilde{\phi}_i|^2)^2 ,
\label{dterm}
\end{equation}
where $\xi$ is a constant and $\widetilde{\phi}_i$ are scalar fields
charged under the anomalous $U(1)$ with charge $q_i$. Such a term, for
vanishing $\widetilde{\phi}_i$, gives a constant term in the potential
but does not contribute in leading order to the masses of uncharged
scalar fields. The latter occur in radiative order only giving a
contribution to their mass-squared equal to $\beta^2\,V_{0}/M^{2}$ where
$\beta^2\approx\,h'^2/16\pi^2$, and typically $\beta\approx10^{-1}$ for
an effective coupling $h'\sim1$ between the charged and neutral
fields.

The other suggested ways out of the $\eta$ problem do not require
non-zero D-terms during inflation but rather provide reasons why the
quadratic term in eq.(\ref{sugra}) should be anomalously small. One
possibility follows from the fact that the mass term of ${\cal O}%
(V_0/M^2)$ coming from eq.(\ref{sugra}) actually applies at the Planck
scale. Since we are considering inflation for field values near the
origin the inflaton mass must be run down to low scales. As shown in
ref.\cite{kingross}, in a wide class of models in which the gauge
couplings become large at the Planck scale the low energy
supersymmetry breaking soft masses are driven much smaller at low
scales by radiative corrections. The typical effect is to reduce the
mass by a factor $\beta\approx\alpha(M)/\alpha(\mu)$ where
$\alpha(\mu)$ is a gauge coupling evaluated at the scale $\mu$. While
radiative corrections can cause a significant change in the coupling,
the effect is limited and becomes smaller as the gauge coupling
becomes small. For this reason the effective mass at low scales cannot
be arbitrarily small and typically $\beta\gtrsim1/25$.

It is also possible to construct models in which the contribution to
the scalar mass exhibited in eq.(\ref{sugra}) is cancelled by further
contributions coming from the expansion of $f$ in eq.(\ref{V}). One
interesting example which arises in specific superstring theories was
discussed in ref.\cite{casas}. Another occurs in supergravity models
of the `no-scale' type \cite{noscale}. In these examples, while the
scalar mass is absent in leading order, it typically arises in
radiative order and so again there is an expectation that the
effective mass cannot be arbitrarily small.

To summarise, in all these cases for small field values the effective
supergravity potential can be conveniently parameterised as a constant
term plus a quadratic term involving a real inflaton field
$\overline{\phi}$ (the real part of $\phi$) i.e. just the form of
eq.(\ref{newinflation}). However this is the form at tree level
only. In general radiative corrections cause the effective mass,
$m_{\overline{\phi}}$ {to depend logarithmically on
$\overline{\phi}$. Thus the full potential has the form
\begin{equation}
 V (\overline{\phi}) = \Delta^4
 \left[1 + b\left(\frac{|\overline{\phi}|}{M}\right)^2
 + c\ln\left(\frac{|\overline{\phi}|}{M}\right)
  \left(\frac{|\overline{\phi}|}{M}\right)^2\right] .
\label{pot}
\end{equation}
Note that in eq.(\ref{pot}) we have not included a term linear in $%
\overline{\phi}$. Such a term can be forbidden if the theory has a
symmetry under which $\overline{\phi}$ transforms non-trivially and we
assume that this is indeed the case.

\subsection{Terminating slow-roll evolution} \label{end}

The form of eq.(\ref{pot}) provides a useful parameterisation of the
inflationary potential in the neighbourhood of the origin. However
when $\overline{\phi}/M$ is large higher order terms may be
significant and so it is important to determine their structure. Such
terms may occur through higher order terms in $d$ or $f$ in
eq.(\ref{g}) or in higher order terms in the expansion of
eq.(\ref{sugra}). Typically these terms are expected be associated
with new physics at the Planck scale, for example a term in the
superpotential of the form $\phi^{p/2+1}/M^{p/2-2}$ will give rise to
terms of the form $|\phi|^{p}/M^{p-4}$ in the scalar
potential. However it may happen that such higher order terms arise as
a result of integrating out heavy fields in the theory, in which case
the mass scale in the denominator can be much smaller than $M$. For
example the superpotential $f=\phi^{2}X+M_{X}X^{2}$ describes a field
$X$ with mass $M_X$. At scales below $M_X$ the field $X$ may be
integrated out to give the effective superpotential $f=\phi^4/M_X$ in
which the scale of the higher dimension operator is set by
$M_X$. This mass may be associated with any of the scales in the
theory, e.g. the inflationary scale or the supersymmetry breaking
scale, and can thus be much smaller than the Planck scale.

With this preamble we now consider the structure of the terms
responsible for ending inflation Consider first the inflationary
models driven by a non-zero F-term. This is conveniently parameterised
by the superpotential $f=\Delta^{2}Y$ which gives $V=|F_Y|^2=\Delta^4$
as required. Radiative corrections then lead to the form of
eq.(\ref{pot}). Now consider the form of higher order corrections to
this superpotential. The terms allowed must be consistent with any
gauge or discrete symmetries of the models. For example the above form
linear in $Y$ follows if $Y$ carries non-zero $R$-symmetry charge
under an unbroken $R$-symmetry. Using such symmetries it is
straightforward to construct a general potential displaying the
possibilities for ending inflation. We suppose that $\phi$ is a
singlet under the $R$-symmetry but carries a charge under a discrete
$Z_{p}$ symmetry. Then the superpotential has the form
\begin{equation}
 f = \left(\Delta^2 - \frac{\phi^p}{M'^{p-2}}
     - \frac{\phi^{2p}}{M'^{2p-2}} - \dots \right)Y
\end{equation}
where we have suppressed the coefficients of ${\cal O}(1)$ of each term.
This gives rise to the potential
\begin{equation}
 V = \left(\Delta^2 - \frac{\phi^p}{M'^{p-2}}
     - \frac{\phi^{2p}}{M'^{2p-2}} - \dots \right)^2 ,
\label{einf}
\end{equation}
plus terms involving $Y$ which we drop as they do not contribute to
the vacuum energy (since $Y$ does not acquire a vacuum expectation
value). This has the desired structure for ending inflation because
the potential vanishes for $\phi\sim\,M'(\Delta/M')^{2/p}$ for
$M'>\Delta$. For our analysis we use a slightly simplified form of
eq.(\ref{einf}) (c.f. eq.(\ref{einfp})) keeping only the leading $\phi
^{p}$ term and setting $M'^{p-2}=\Delta^{q}M^{p-q-2}$ to take account
of the possibility discussed above that the scale associated with the
higher dimension operators may be below the Planck scale.

The situation is similar for the case of $D$-term inflation. From
eq.(\ref{dterm}) we see there is a contribution of the form of
eq.(\ref{einf}). If $\widetilde{\phi_i}$ is a heavy field it may be
integrated out in a similar manner to that discussed above so that the
term $\sum_{i}q_i|\widetilde{\phi}_i|^2$ may give rise to a higher
dimension term of the form $\phi^p/M'^{p-2}.$ Thus the
parameterisation of eq.(\ref{einfp}) may apply to this case as well.

\section{Quadratic supergravity inflation} \label{para} }

We wish to study locally supersymmetric slow-roll inflationary models
in which the inflaton field evolves from small to large field
values. Near the origin the term involving the lowest power of the
inflaton field is the most important in the potential, hence a simple
parameterisation is adequate. As discussed above, we may take the
inflaton field, $\overline{\phi}$, to be real with a quadratic
potential of the form (\ref{pot}). It is convenient to work with
dimensionless quantities
\begin{equation}
 \tilde{V} (\tilde{\phi}) \equiv \frac{V(\overline{\phi}/M)}{\Delta^4}\ ,
 \qquad
 \tilde{\phi} \equiv \frac{\overline{\phi}}{M}\ ,
 \qquad
 \tilde{\Delta} \equiv \frac{\Delta}{M}\ ,
\end{equation}
in terms of which the potential and its derivatives are given by
\begin{eqnarray}
 \tilde{V} (\tilde{\phi}) &=& 1 + b\tilde{\phi}^2
  + c \tilde{\phi}^2 \ln\tilde{\phi}^2 , \nonumber \\
 \tilde{V}' (\tilde{\phi}) &=& 2\tilde{\phi} (b + c
  + c\ln\tilde{\phi}^2) , \nonumber \\
 \tilde{V}'' (\tilde{\phi}) &=& 2 (b + 3c + c\ln\tilde{\phi}^2).
\label{VV}
\end{eqnarray}
During slow-roll inflation with $\tilde{\phi}\ll1$, one has
\begin{equation}
 \tilde{V} (\tilde{\phi}) \approx 1, \qquad \epsilon \ll |\eta| ,
\end{equation}
hence at $\tilde{\phi}=\tilde{\phi}_\H$, the spectrum of the scalar
perturbations (\ref{deltah}) can be written as
\begin{equation}
 \delta_\H \simeq \pm \frac{2\tilde{\Delta}^2}{\sqrt{75}\pi(1 - n_\H + 8c)
  \tilde{\phi}_\H}\ ,
\label{delta}
\end{equation}
where the spectral index is
\begin{equation}
 n_\H \simeq 1 + 2 \tilde{V}'' (\tilde{\phi}_\H)
      = 1 + 4b + 12c + 8c\ln\tilde{\phi}_\H .
\label{si}
\end{equation}
The $+(-)$ sign in eq.(\ref{delta}) takes into account that
$\tilde{V}'$ is positive (negative) at $\tilde{\phi}=\tilde{\phi}_\H$
corresponding to the inflaton rolling towards smaller (bigger) values
of $\tilde{\phi}$. We focus on the latter case as it offers a simple
way of ending inflation as the field value increases and higher order
terms in $\tilde{\phi}$ become important.

The number of e-folds before the end of inflation when the
fluctuations observed by COBE were generated is
\begin{equation}
 N_\COBE \simeq 48 + \ln\left(\frac{\Delta}{10^{10}\GeV}\right) ,
\label{Ncobe}
\end{equation}
assuming instant reheating; the numerical value would be smaller if
reheating is inefficient or if there are later episodes of `thermal
inflation' \cite{review}. In our model the end of inflation occurs due
to terms of higher order in $\phi$ of the form given in
eq.(\ref{einf}). For the moment we parameterise their effect through
the inclusion of a further parameter $\phi_\e$, the value of the field
at the end of inflation. The number of e-folds from $\phi_\H$ onwards
should equal or exceed $N_\COBE$, giving
\begin{equation}
 N_\H \equiv -\int_{\tilde{\phi}_\H}^{\tilde{\phi}_\e}
         \frac{V(\tilde{\phi})}{V'(\tilde{\phi})} \df\tilde{\phi}
      = \frac{1}{4c}
 \ln\left(\frac{b+c+2c\ln\tilde{\phi}_\H}{b+c+2c\ln\tilde{\phi}_\e}\right)\geq
  N_{\COBE} .
\label{efolds}
\end{equation}
In addition the slow-roll conditions during inflation require
\begin{equation}
 |V''(\tilde{\phi})| \ll \gamma ,\quad {\rm for} \quad
 \tilde{\phi}_\H < \tilde{\phi} < \tilde{\phi}_\e ,
\label{vpp}
\end{equation}
where $\gamma$ is of ${\cal O}(1)$ (and can be calculated exactly in the
slow-roll formalism).

If eqs.(\ref{delta}, \ref{si}, \ref{efolds}) satisfy the observational
constraints (\ref{cobe},\ref{ncobe}, \ref{Ncobe}) the inflationary era
will lead to a satisfactory cosmology. For $b\lesssim25c$ the term
proportional to $b$ in eq.(\ref{pot}) is negligible. We shall
concentrate on this limit (although acceptable inflation is also
possible if this is not the case). Then eq.(\ref{si}) constrains the
quantity
$\case{1}{c}\ln[\ln(\tilde{\phi}_H)/\ln(\tilde{\phi}_e)]$. Further,
eq.(\ref{delta}) constrains a combination of $\tilde{\Delta}$, $c$,
and $\tilde{\phi}_\H$.  Given that we have four parameters to satisfy
these equations and that eq.(\ref{vpp}) can be satisfied for a range
of parameters, we see that acceptable inflation follows for a range of
$\Delta$. This opens up the possibility of having a much lower value
for the inflationary energy scale than is usually considered. The
value of $\Delta$ is very sensitive to the the value of
$\tilde{\phi}_e$ and so it is necessary to consider in some detail
what determines the latter. We will discuss this shortly but first we
consider how a reduction in the scale of inflation, $\Delta$, may
allow thermal effects to naturally set the initial conditions for
inflation.

\subsection{The begining of inflation}

We assume that some process at the Planck scale, presumably quantum
cosmological in nature \cite{hawking}, creates a homogeneous patch of
space-time and releases energy thermalising the universe. It is
important to note that the thermalisation temperature cannot be close
to the Planck scale, regardless of the amount of energy released,
essentially because particle interactions are asymptotically free. For
example, an explicit calculation of the $q\bar{q}$ annihilation rate
into gluons finds that equilibrium can only be attained below a
temperature $T_\therm\sim3\times10^{14}$~GeV
\cite{enqvist1}.\footnote{A similar estimate of the thermalisation
temperature obtains in a study where cold particles are released at
the Planck scale and allowed to scatter to achieve equilibrium
\cite{enqvist2}.}

Let us consider the requirements on the inflaton field for it to be
localised at the origin through its couplings to particles in the
thermal bath. On dimensional grounds, the $2\to2$
scattering/annihilation cross-sections at energies higher than the
masses of the particles involved are expected to decrease with
increasing temperature as $\sim\alpha^2/T^2$, where $\alpha$ is the
coupling. Thus if the scattering/annihilation rate,
$\Gamma\sim\,n\langle\sigma\,v\rangle$ is to exceed the Hubble
expansion rate $H_\therm\sim(g_*T^4/10M^2)^{1/2}$ in the
radiation-dominated plasma, then we have a limit on the thermalisation
temperature $T_\therm\lesssim\alpha^2M/3g_*^{1/2}$, where $g_*$ counts
the relativistic degrees of freedom (=915/4 in the minimal
supersymmetric standard model (MSSM) at high temperatures). Now a
Yukawa coupling $h'\phi\chi^2$ of the inflaton to MSSM fields $\chi$
will generate a confining potential at high temperatures,
$V(\phi,T)\sim\,h'^2T^2\phi^2$ i.e. an effective mass for the inflaton
of $m_\therm\sim\,h'T$. This will rapidly drive the inflaton field to
the origin in a time of ${\cal O}(m_\therm^{-1})$ As the universe
cools, the potential energy $\sim\Delta^4$ of the inflaton will begin
to dominate over the thermal energy at a temperature
$T_\inf\sim\Delta^2/\alpha^2M$. At this epoch the inflaton field will
be localised to a region $\delta\phi\sim\,T_\inf$ in the neighbourhood
of the origin. Thus to provide natural initial conditions for new
inflation we require that $T_\inf<T_\therm$ i.e. that the scale of
inflation be sufficiently low. The discussion above yields the
constraint $\Delta\lesssim10^{-4}M$, taking $\alpha\sim1/24$.

\subsection{The end of inflation}

As mentioned earlier, the end of inflation will be determined by terms
which are {\em not} included in the expansion of the inflationary
potential keeping only terms up to quadratic order. In general such
terms are quite model dependent as all orders in $\phi$ may contribute
significantly for large $\phi$. However it is necessary to discuss the
expectation for $\phi_\e$ for reasonable forms of the potential since
the nature of the inflationary era is strongly dependent on it. In
Section~\ref{eta} we have discussed how higher order terms arise in
supergravity theories and how they are restricted by the symmetries of
the theory but here we will simply write down the structure of these
terms motivated by such considerations. Although this parameterisation
involves only a single inflaton field in the present context , it
actually covers a wide range of models including hybrid inflationary
models in which the end of inflation is triggered by a second
field. The full inflationary potential then has the form:
\begin{equation}
 V (\phi) = \left(1 - \kappa\frac{\tilde{\phi}^p}{\Delta^q}\right)^2
            + b\tilde{\phi}^2 + c\tilde{\phi}^2\ln\tilde{\phi}^2 ,
\label{einfp}
\end{equation}
where $\kappa$ is a coupling which for simplicity we set to unity
(its most natural value?) in our calculations.

\begin{figure}[h]
\hspace{2cm}\epsfxsize12cm\epsffile{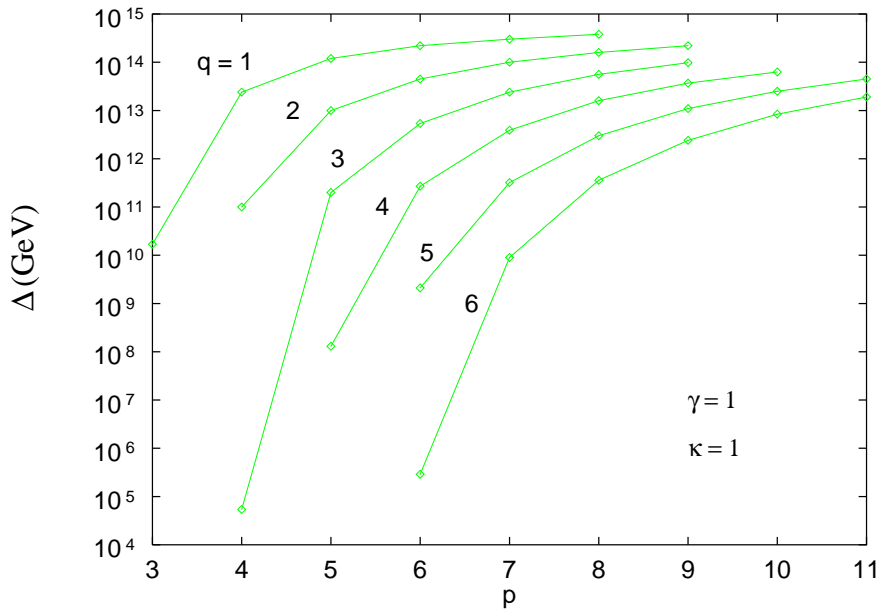}
\caption{The inflationary energy scale for various choices of the
non-renormalisable term which terminates inflation.}
\label{fig1}
\end{figure}

The end of slow-roll occurs at
\begin{equation}
 \tilde{\phi}_\e =
 \left(\frac{\gamma\Delta^q}{2\kappa p(p - 1)}\right)^{1/(p-2)} .
\label{phie}
\end{equation}
Using this we may now determine the range of parameters for which
acceptable inflation obtains by identifying $\tilde{\phi}_\H$ with the
field value at which the fluctuations observed by COBE were
generated. In Figure~\ref{fig1} we show the required inflationary
scale for various values of $p, q$, setting $\gamma, \kappa=1$ and
adopting $c=b/25$. The essential observation is that the quadratic
potential generates acceptable inflation for a wide range of values of
$\Delta$. This is important because it opens up the possibility of
identifying the inflationary scale with a mass scale already required
in the particle physics model. Further if $\Delta\lesssim10^{14}\GeV$,
thermal effects will be important in setting the initial conditions
for inflation. For the cases shown in Figure~\ref{fig1}, inflation
typically ends at $\tilde{\phi}_\e\sim10^{-8}$ and the fluctuations
observed by COBE exited the Hubble radius at
$\tilde{\phi}_\COBE\sim10^{-11}-10^{-9}$, about 50 e-folds of
expansion earlier. We have checked that in all cases
$\tilde{\phi}_\COBE$ exceeds $T_\inf$ so that there will be sufficient
inflation starting with thermal initial conditions.

\section{Conclusions}

We have studied in very general terms, an inflationary era driven by a
quadratic potential. The essential requirement for successful
inflation is that the inflaton mass be reduced below the Hubble
parameter. We have discussed mechanisms which do this, concentrating
on the situation where the inflaton mass-squared is negative at the
origin.\footnote{Other studies of quadratic inflation have
concentrated on the case where radiative corrections make the
potential develop a maximum near the origin, from which the inflaton
rolls either towards the origin \cite{stewart} or away from it
\cite{covi}, and inflation ends through a hybrid mechanism.} This has
the advantage that thermal initial conditions naturally place the
inflaton at the origin, initiating the inflationary era. Whether such
inflation occurs is a model dependent question. In compactified string
theories the multiplet structure is completely determined and
typically there are many scalar fields with the properties of the
inflaton discussed here. In such models inflation driven by a
quadratic potential is a very natural possibility.

In order for the above mechanism to work, the inflaton must couple to
the fields in the thermal bath and the scale of its potential must be
sufficiently low. In fact the quadratic parametrization of the
inflationary potential allows any value for the inflationary scale
$\Delta$. This is essentially because the end of inflation occurs not
due to the violation of the slow-roll conditions by the quadratic
potential, but due to higher order non-renormalisable terms which
become important as the inflaton evolves towards large field values.
Two values for the inflationary scale are of particular interest and
will be discussed in detail elsewhere \cite{progress}. One has $%
\Delta\sim10^{10}~\GeV$, the supersymmetry breaking scale in the
hidden sector. The other with $\Delta\sim10^3~\GeV$ implements
inflation at the electroweak scale and could be relevant in the
context of theories with submillimeter dimensions.

\acknowledgements

G.G. acknowledges financial support by UNAM and CONACyT, M\'exico.


\begin{references}

\bibitem{book}
 see, A.D. Linde, {\sl Particle Physics and Inflationary Cosmology}
  (Harwood Academic Press, 1990).

\bibitem{review}
 For a review and extensive references, see, D.H. Lyth and A. Riotto,
  Phys. Rep. 314 (1999) 1.

\bibitem{chaotic}
 A.D. Linde, Phys. Lett. 129B (1983) 177.

\bibitem{cobe}
 C.L. Bennett {\em et al} (COBE collab.), Astrophys. J. 464 (1996) L1.

\bibitem{new}
 A.D. Linde, Phys. Lett. 108B (1982) 389;
 A. Albrecht and P.J. Steinhardt, Phys. Rev. Lett. 48 (1982) 1220.

\bibitem{susybreak}
 M. Dine, W. Fischler and D. Nemechansky, Phys. Lett. 136B (1984) 169;
 G. Coughlan, W. Fischler, E.W. Kolb, S. Raby and G.G. Ross,
  Phys. Lett. 140B (1984) 44.

\bibitem{copeland}
 E. Copeland, A.R. Liddle, D.H. Lyth, E.D. Stewart and D. Wands,
  Phys. Rev. D49 (1994) 6410.

\bibitem{hawking}
 S.W. Hawking and N. Turok, Phys. Lett. B425 (1998) 25;
 A. Vilenkin, Phys. Rev. D57 (1998) 7069;
 A.D. Linde, Phys. Rev. D58 (1998) 083514
 N. Turok and S.W. Hawking, Phys. Lett. B432 (1998) 271.

\bibitem{largedim}
 N. Arkani-Hamed, S. Dimopoulos and G. Dvali, Phys. Lett. B429 (1998) 263.

\bibitem{bailin}
 D. Bailin and A. Love, {\sl Supersymmetric Gauge Field Theory and String
  Theory} (Adam Hilger, 1994).

\bibitem{enqvist1}
 K. Enqvist and J. Sirkka, Phys. Lett. B314 (1993) 298;

\bibitem{enqvist2}
 K. Enqvist and K.J. Eskola, Mod. Phys. Lett. A5 (1990) 1919.

\bibitem{dterm}
 E.D. Stewart, Phys. Rev. D51 (1995) 6847;
 E. Halyo, Phys. Lett. B387 (1996) 43;
 P. Bin\'{e}truy and G. Dvali, Phys. Lett. B388 (1996) 241;
 for a review and further references, see \cite{review}.

\bibitem{casas}
 J.A. Casas and G.B. Gelmini, Phys. Lett. B410 (1997) 36.

\bibitem{noscale}
 M.K. Gaillard, H. Murayama and K.A. Olive, Phys. Lett. B355 (1995) 71;
 M. Bastero-Gil and S.F. King, Nucl.Phys. B549 (1999) 391.

\bibitem{kingross}
 S.F. King and G.G. Ross, Nucl. Phys. B530 (1998) 3.

\bibitem{stewart}
 E.D. Stewart, Phys. Rev. D56 (1997) 2019.

\bibitem{covi}
 E.D. Stewart, Phys. Lett. B391 (1997) 34;
 L. Covi, D.H. Lyth and L. Roszkowski, Phys. Rev. D60 (1999) 023509;
 L. Covi and D.H. Lyth,  Phys. Rev. D60 (1999) 063515

\bibitem{progress}
 G. Germ\'{a}n, G.G. Ross and S. Sarkar, in preparation.

\end{references}
\end{document}